\documentclass[%
 reprint,
superscriptaddress,
 amsmath,amssymb,
 aps,
prb,
]{revtex4-2}

\usepackage{graphicx}
\usepackage{dcolumn}
\usepackage{bm}
\usepackage[colorlinks=true,urlcolor=magenta,linkcolor=magenta,citecolor=cyan]{hyperref}
\usepackage{verbatim}
\usepackage[utf8]{inputenc}
\usepackage{layouts}
\usepackage{xcolor}
\usepackage{multirow}
\usepackage{hhline} 
\usepackage[separate-uncertainty=true]{siunitx}
\usepackage[version=3]{mhchem} 
\usepackage{layouts}
\usepackage{leftidx} 
\usepackage{amsmath,mathtools}
\usepackage{chngcntr}

\begin{document}


\title[\texttt{achemso} demonstration]
{Raman spectrum of Janus transition metal dichalcogenide monolayers\\WSSe and MoSSe}


\author{Marko M. Petri\'{c}}
\thanks{These authors contributed equally to this work.}
\affiliation{Walter Schottky Institut and Department of Electrical and Computer Engineering, Technische Universit\"{a}t M\"{u}nchen, Am Coulombwall 4, 85748 Garching, Germany}
\affiliation{Munich Center for Quantum Science and Technology (MCQST), Schellingstrasse 4, 80799 Munich, Germany}

\author{Malte Kremser}
\thanks{These authors contributed equally to this work.}
\affiliation{Walter Schottky Institut and Physik-Department, Technische Universit\"{a}t M\"{u}nchen, Am Coulombwall 4, 85748 Garching, Germany}
\affiliation{Munich Center for Quantum Science and Technology (MCQST), Schellingstrasse 4, 80799 Munich, Germany}

\author{Matteo Barbone}
\thanks{These authors contributed equally to this work.}
\affiliation{Walter Schottky Institut and Department of Electrical and Computer Engineering, Technische Universit\"{a}t M\"{u}nchen, Am Coulombwall 4, 85748 Garching, Germany}
\affiliation{Munich Center for Quantum Science and Technology (MCQST), Schellingstrasse 4, 80799 Munich, Germany}

\author{Ying Qin}
\affiliation{Materials Science and Engineering, School for Engineering of Matter, Transport and Energy, Arizona State University, Tempe, Arizona 85287, USA}

\author{Yasir Sayyad}
\affiliation{Materials Science and Engineering, School for Engineering of Matter, Transport and Energy, Arizona State University, Tempe, Arizona 85287, USA}

\author{Yuxia Shen}
\affiliation{Materials Science and Engineering, School for Engineering of Matter, Transport and Energy, Arizona State University, Tempe, Arizona 85287, USA}

\author{Sefaattin Tongay}
\email{Sefaattin.Tongay@asu.edu}
\affiliation{Materials Science and Engineering, School for Engineering of Matter, Transport and Energy, Arizona State University, Tempe, Arizona 85287, USA}

\author{Jonathan J. Finley}
\email{finley@wsi.tum.de}
\affiliation{Walter Schottky Institut and Physik-Department, Technische Universit\"{a}t M\"{u}nchen, Am Coulombwall 4, 85748 Garching, Germany}
\affiliation{Munich Center for Quantum Science and Technology (MCQST), Schellingstrasse 4, 80799 Munich, Germany}

\author{Andr\'{e}s R. Botello-M\'{e}ndez}
\email{botello@fisica.unam.mx}
\affiliation{Universidad Nacional Aut\'{o}noma de M\'{e}xico, Institute of Physics, 20-364, 01000 M\'{e}xico, D.F., M\'{e}xico}

\author{Kai M\"{u}ller}
\email{Kai.Mueller@wsi.tum.de}
\affiliation{Walter Schottky Institut and Department of Electrical and Computer Engineering, Technische Universit\"{a}t M\"{u}nchen, Am Coulombwall 4, 85748 Garching, Germany}
\affiliation{Munich Center for Quantum Science and Technology (MCQST), Schellingstrasse 4, 80799 Munich, Germany}

\begin{abstract}

Janus transition metal dichalcogenides (TMDs) lose the horizontal mirror symmetry of ordinary TMDs, leading to the emergence of additional features, such as native piezoelectricity, Rashba effect, and enhanced catalytic activity. While Raman spectroscopy is an essential nondestructive, phase- and composition-sensitive tool to monitor the synthesis of materials, a comprehensive study of the Raman spectrum of Janus monolayers is still missing. Here, we discuss the Raman spectra of WSSe and MoSSe measured at room and cryogenic temperatures, near and off resonance. By combining polarization-resolved Raman data with calculations of the phonon dispersion  and using symmetry considerations, we identify the four first-order Raman modes and higher-order two-phonon modes. Moreover, we observe defect-activated phonon processes, which provide a route toward a quantitative assessment of the defect concentration and, thus, the crystal quality of the materials. Our work establishes a solid background for future research on material synthesis, study, and application of Janus TMD monolayers.

\end{abstract}

\maketitle


\section{Introduction}
Over the past decade, transition metal dichalcogenide (TMD) monolayers have emerged as a unique playground for exciton photophysics due to $0.5$-eV-high exciton binding energies \cite{Chernikov.2014, Wang.2018}, strong light-matter interaction \cite{Bernardi.2013, Wurstbauer.2017}, optically addressable valley-contrasting spin physics caused by broken inversion symmetry \cite{DiXiao.2012}, and large spin-orbit coupling \cite{Zhu.2011}. Further studies also revealed the presence of quantum light emitters \cite{Tonndorf.2015, Koperski.2015, Srivastava.2015, Chakraborty.2015, He.2015, PalaciosBerraquero.2016} and evidence of strongly correlated phases such as superconductivity \cite{Ye.2012, Costanzo.2016} and exciton and polaron Bose-Einstein condensation \cite{Cotlet.2016, Kogar.2017}. Integrability on conventional silicon photonic technology \cite{Tonndorf.2017, Youngblood.2017}, large-area fabrication \cite{Lee.2012, Zhan.2012, Lee.2013} and deterministic positioning of quantum emitters \cite{PalaciosBerraquero.2017, Branny.2017, Klein.2019} widen the impact of TMDs to optoelectronics \cite{Ross.2014} and energy harvesting \cite{Pospischil.2014} as well as applications exploiting valleytronics \cite{Jones.2013, Xu.2014}, spintronics \cite{Mak.2014, Yan.2016, Han.2016}, and quantum photonic properties \cite{Atature.2018, Aharonovich.2016}. In addition, new functionalities and physical phenomena appear when stacking monolayers of TMDs on top of each other, forming  artificial metamaterials held together by van der Waals forces \cite{Geim.2013}. Such heterostructures host long-lived, tunable dipolar interlayer \cite{Rivera.2015} and trapped moir\'{e} excitons \cite{Seyler.2019, Jin.2019, Tran.2019, Alexeev.2019}, offering a rich playground for few- and many-body phenomena \cite{Mak.2013, Barbone.2018, Li.2018, Ye.2018, Chen.2018, Hao.2017, Kremser.2020}, making them candidates for a solid-state quantum simulation platform \cite{Wu.2018}.

In contrast to conventional, mirror-symmetric TMDs with a stoichiometric formula $MX_2$ (where $M$ is a transition metal and $X$ is a chalcogen), Janus TMD monolayers $MXY$ are formed when the crystal plane of transition metal atoms is sandwiched between two planes, each made of a different chalcogen atom $X$ and $Y$. This breaks mirror symmetry along the direction perpendicular to the plane of the two-dimensional (2D) material, reducing the overall symmetry of the crystal, and gives rise to an intrinsic electrical dipole in the unit cell created by the difference in electronegativity between the top and bottom chalcogen atoms \cite{Li.2017}. Consequently, theoretical studies predict the appearance of a multitude of physical phenomena such as piezoelectricity \cite{Dong.2017, Guo.2017}, enhanced photocatalysis \cite{Er.2018, Guan.2018, Ji.2018, Xia.2018}, Rashba splitting \cite{Cheng.2013, Hu.2018, Yao.2017}, and the presence of topological phases \cite{Ma.2018}. However, most of these physical effects are still experimentally unexplored, due to the very recent success
at synthesizing Janus crystals. First reports on the growth of Janus MoSSe appeared in 2017 \cite{Lu.2017, Zhang.2017}, followed by Janus WSSe \cite{Trivedi.2020,Lin.2020}, and their heterostructures in 2020 \cite{Trivedi.2020}.

Inelastic light scattering is a powerful, non-destructive tool to gain insight into the structural and electronic properties of materials \cite{Zhang.2015, Saito.2016}. Each Raman spectrum of 2D materials is a unique fingerprint of a sample, shedding light on its crystal and electronic band structure \cite{Sun.2013}, layer number \cite{Li.2012}, interlayer coupling \cite{Zhao.2013, Verble.1970}, doping \cite{Chakraborty.2012}, defect density \cite{Mignuzzi.2015}, electron-phonon interaction \cite{Carvalho.2015}, etc. Moreover, Raman spectroscopy can be used \textit{in situ} during growth to distinguish a Janus monolayer from a disordered ternary alloy \cite{Komsa.2012, Su.2014, Mann.2014, Li.2015, Taghizadeh.2020}. Thus, to synthesize high-quality Janus TMD crystals and unlock the predicted effects and applications it is crucial to have a detailed study of its vibrational spectrum. However, a comprehensive study of the Raman spectrum of Janus monolayers WSSe and MoSSe is lacking. Initial experimental measurements have given limited insight on the Raman spectrum of Janus TMD monolayers, suffering from incomplete \cite{Lu.2017} or even incorrect \cite{Zhang.2017} assignments of the first-order modes and leaving all other features unidentified.

In this work, we calculate the phonon band structure of Janus WSSe and MoSSe monolayers and their phonon density of states (PhDOS), which we use to predict the Raman modes and their energies. Then, we measure the Raman spectra of both materials at room and cryogenic (10 K) temperature and two excitation wavelengths $\lambda_{\mathrm{ex}}$, closer to and farther from excitonic resonances. Further, we perform polarization-resolved Raman measurements at room temperature and $\lambda_{\mathrm{ex}}=\SI{532}{\nano\meter}$. By comparing theory and experiments, we identify the first-order Raman modes. As the experimental spectra show rich features arising beyond the calculated first-order processes, we then discuss the mechanisms of higher-order and defect-mediated Raman modes and assign them to the relevant experimental peaks.


\section{Results and Discussion}

Monolayers of conventional 2\textit{H}-TMDs have $D_\mathrm{3h}^1$ space group symmetry \cite{sandoval1991raman, RibeiroSoares.2014}, contrasting strongly with Janus TMD monolayers, for which rotation $C_2$, improper rotation $S_3$, and mirror $\sigma_{\mathrm{h}}$ symmetries are broken due to the different chalcogen atoms in the unit cell. This results in a lowering of the symmetry of the crystal to the symmorphic (i.e., all symmetry operations leave one common point fixed) $C_\mathrm{3v}^1$ space group ($C_{\mathrm{3v}}$ point group). The unit cell of the Janus monolayer $MXY$ is formed from three atoms, resulting in $3\times3=9$ normal vibrational modes at the $\mathbf{\Gamma}$ point (center) of the Brillouin zone, of which three are acoustic and six are optical. Group theory identifies these vibrations as the irreducible representations of the $C_{\mathrm{3v}}$ point group, that can be expressed by $\Gamma_{C_{\mathrm{3v}}}^{\mathrm{vib}} = 3A_1 (\Gamma_1) + 3E(\Gamma_3) $, where all of the modes are both Raman and infrared (IR) active. Here, $\Gamma_{C_{\mathrm{3v}}}^{\mathrm{vib}}$ is the irreducible representation of the total vibration, deduced from the underlying crystal symmetry using the $C_{\mathrm{3v}}$ character table. In-plane vibrations are defined as $E^{1,2}$ and out-of-plane as $A_1^{1,2}$, with $E^{1,2}$ being doubly degenerate at the $\mathbf{\Gamma}$ point (modes with the same symmetry are distinguished by the upper right corner index).

\begin{figure*}[ht!]
\centering
\includegraphics{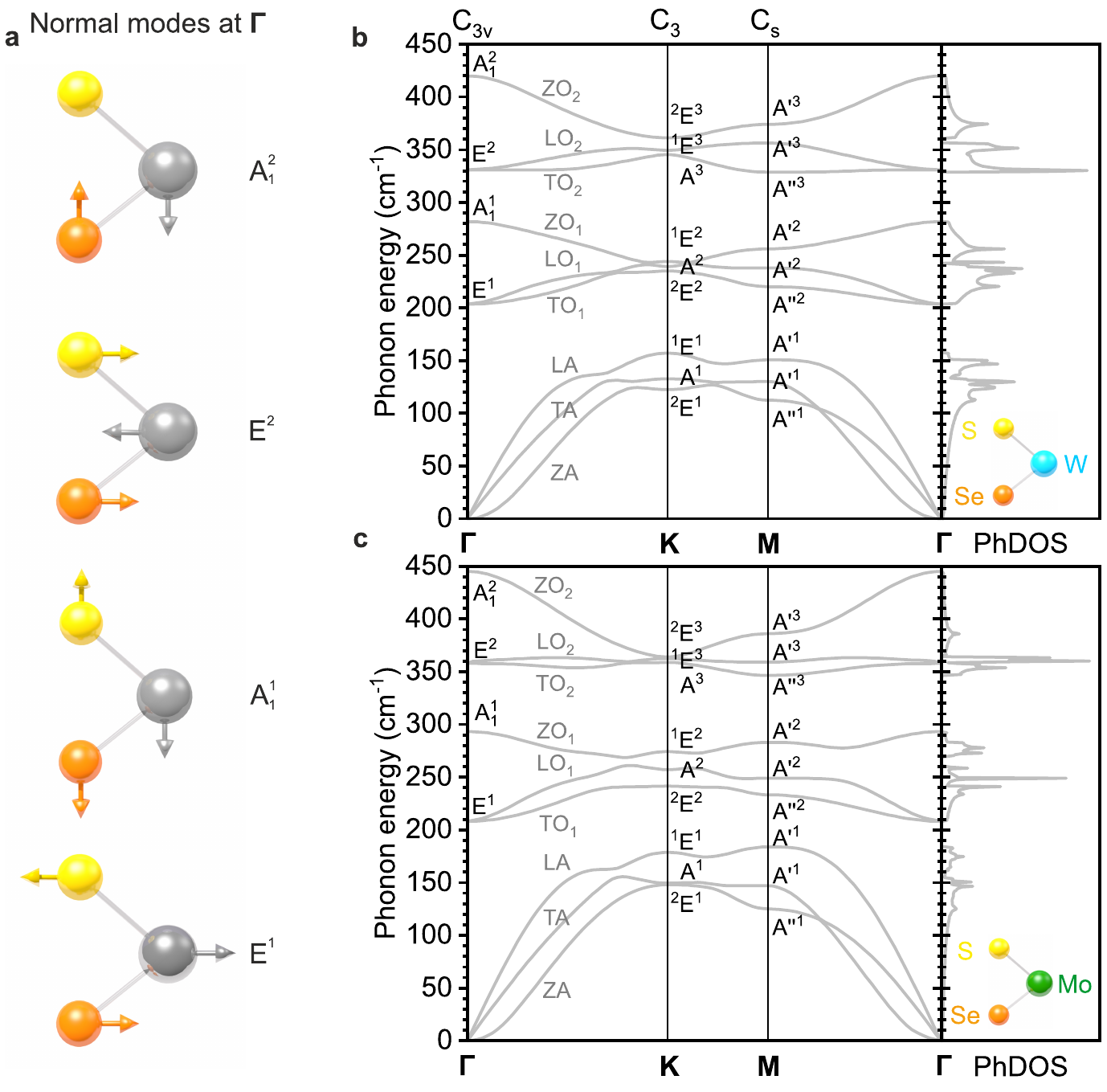}
\caption{\textbf{First-order phonons and phonon band structure of Janus TMD monolayers WSSe and MoSSe.} \textbf{(a)} Schematic representation of the atomic vibrations at the center $\mathbf{\Gamma}$ of the Brillouin Zone. Transition metal, selenium and sulfur atoms are identified by gray, orange, and yellow, respectively. DFPT calculations of the phonon band structure of WSSe [\textbf{(b)}, left panel] and MoSSe [\textbf{(c)}, left panel]. Each energy dispersion diagram shows three acoustic (ZA, TA, LA) and six optical ($\mathrm{ZO_1}$, $\mathrm{ZO_2}$, $\mathrm{TO_1}$, $\mathrm{TO_2}$, $\mathrm{LO_1}$, $\mathrm{LO_2}$) phonon branches. The corresponding PhDOS  are shown in \textbf{(b)} and \textbf{(c)} (right panels). Vibrational modes are labeled at the high-symmetry points $\mathbf{\Gamma}$ ($E^{1,2}$, $A_1^{1,2}$), \textbf{K} ($\prescript{1}{}E^{1,2,3}$, $\prescript{2}{}E^{1,2,3}$, $A^{1,2,3}$), and \textbf{M} ($A'^{1,2,3}$, $A''^{1,2,3}$).}
\label{fig:Figure1}
\end{figure*}

Owing to the conservation of energy and quasimomentum $\mathbf{q}$ in the crystal, first-order (i.e., one-phonon) scattering processes are bound to the $\mathbf{\Gamma}$ point of the Brillouin zone due to the negligible photon momentum ($\mathbf{q}\approx\mathbf{0}$). The atomic displacements corresponding to the normal vibrational modes at $\mathbf{\Gamma}$ are schematically represented in Fig. \hyperref[fig:Figure1]{1(a)} with the transition metal atom in gray and the chalcogens Se and S in orange and yellow, respectively. We used density functional perturbation theory (DFPT) to predict the phonon modes (see Appendix \hyperref[sec:A]{A}). In WSSe at the $\mathbf{\Gamma}$ point, they occur at 204, 282, 331, and \SI{420}{\centi\meter^{-1}} (for $E^1$, $A_1^1$, $E^2$, and $A_1^2$, respectively). Analogously, in MoSSe at the $\mathbf{\Gamma}$ point, they occur at 208, 293, 358, and \SI{445}{\centi\meter^{-1}} (for $E^1$, $A_1^1$, $E^2$, and $A_1^2$, respectively). This can be seen in the phonon band structure of monolayer WSSe in Fig. \hyperref[fig:Figure1]{1(b)} and of monolayer MoSSe in Fig. \hyperref[fig:Figure1]{1(c)}. The three acoustic phonon branches correspond to the out-of-plane acoustic (ZA), the transverse acoustic (TA), and the in-plane longitudinal acoustic (LA) modes, respectively. The remaining six branches represent the out-of-plane optical (ZO$_1$ and ZO$_2$), the in-plane transverse optical (TO$_1$ and TO$_2$), and the in-plane longitudinal optical (LO$_1$ and LO$_2$) modes.

In addition to the $\mathbf{\Gamma}$ point, we further examine the vibrational modes at high-symmetry points at the Brillouin zone edge, \textbf{K} and \textbf{M} (see Appendix \hyperref[sec:B]{B}). At the \textbf{K} point, the crystal exhibits $C_3$ point group symmetry with the irreducible representation $\Gamma_{C_3}^{\mathrm{vib}} = 3A (K_1) + 3\prescript{2}{}E (K_2) + 3\prescript{1}{}E (K_3) $, where $\prescript{1}{}E^{1,2,3}$, $\prescript{2}{}E^{1,2,3}$, and $A^{1,2,3}$ are all Raman active modes. On the other hand, at the \textbf{M} point, the crystal exhibits $C_\mathrm{s}$ point group symmetry with the irreducible representation $\Gamma_{C_{\mathrm{s}}}^{\mathrm{vib}} = 6A'(M_1) +  3A''(M_2) $, where $A'^{1,2,3}$ and $A''^{1,2,3}$ are both Raman active modes. Accompanied to the phonon dispersion in Figs. \hyperref[fig:Figure1]{1(b)} and \hyperref[fig:Figure1]{1(c)}, the phonon density of states (PhDOS) reveals a high density of phonons at the flat bands, in particular close to the high-symmetry points \textbf{K} and \textbf{M}, with all phonons being Raman active. The dispersion branches of WSSe in Fig. \hyperref[fig:Figure1]{1(b)} are energetically lower than the dispersion branches of MoSSe in Fig. \hyperref[fig:Figure1]{1(c)}, thus giving lower phonon energies at the same point in the Brillouin zone. This mainly occurs due to the larger atomic mass of W, which makes the vibrations softer, as in the case of regular Mo- and W-based TMDs \cite{MolinaSanchez.2011}. The two materials also differ in the values of the phonon bandgaps (see Appendix \hyperref[sec:C]{C}).

We measured the Raman spectra of Janus TMD monolayers recorded from crystals grown via room temperature selective epitaxy atomic replacement (SEAR) \cite{Trivedi.2020}, as described in Appendix \hyperref[sec:A]{A}. Here, the top-layer selenium atoms, in already grown $\mathrm{WSe_2}$ and $\mathrm{MoSe_2}$ monolayers, are replaced by sulfur atoms, to eventually yield Janus TMD WSSe and MoSSe, respectively. WSSe was grown on $\mathrm{Al_2O_3}$, whereas MoSSe was grown on a $\mathrm{Si/SiO_2}$ substrate. We first conducted Raman spectroscopy in a back-scattering configuration, with a linearly polarized excitation and no polarization filtering of the Raman signal.

\begin{figure*}[ht!]
\centering
\includegraphics{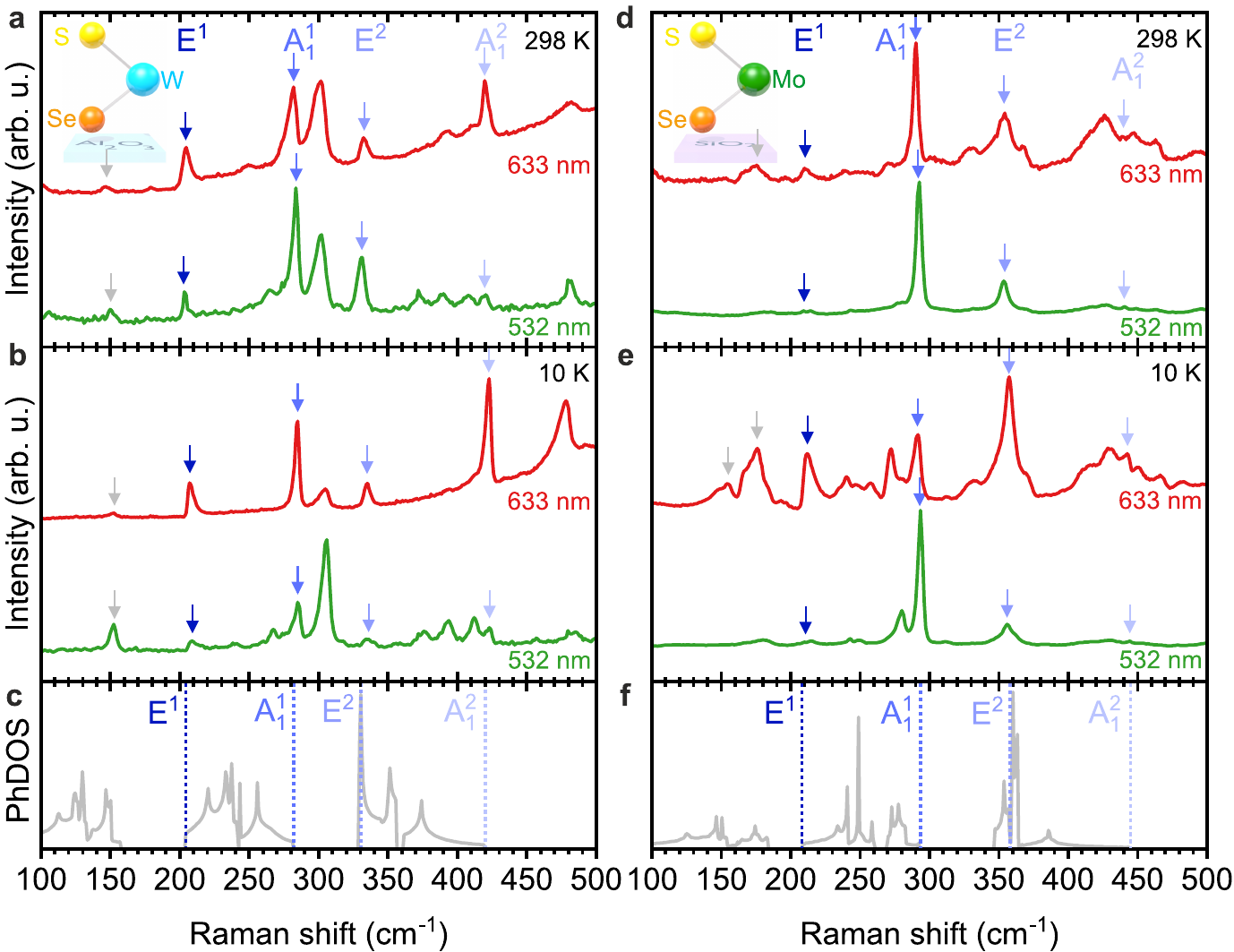}
\caption{\textbf{Raman spectra of Janus TMD monolayers and first-order phonon modes.} \textbf{(a)} Room temperature and \textbf{(b)} 10-K Raman spectra of WSSe at $\lambda_{\mathrm{ex}}=\SI{532}{\nano\meter}$ (green curves) and $\lambda_{\mathrm{ex}}=\SI{633}{\nano\meter}$ (red curves). Blue and gray arrows indicate first-order and defect-activated Raman modes, respectively. \textbf{(c)} PhDOS of WSSe, with the calculated positions of the first-order Raman modes identified by dashed blue lines. \textbf{(d)},\textbf{(e)} Corresponding Raman spectra and \textbf{(f)} PhDOS with predicted first-order Raman modes of MoSSe.}
\label{fig:Figure2}
\end{figure*}

Figure \ref{fig:Figure2} shows the Raman spectra of Janus monolayers WSSe and MoSSe between 100 and \SI{500}{\centi\meter^{-1}}, collected with a laser excitation wavelength $\lambda_{\mathrm{ex}}$ of \SI{532}{\nano\meter} (green curves) and \SI{633}{\nano\meter} (red curves), above the excitonic band gap at \SI{10}{\kelvin} of both WSSe \cite{Trivedi.2020} ($\sim$\SI{670}{\nano\meter}) and MoSSe \cite{Lu.2017, Zhang.2017, Trivedi.2020} ($\sim$\SI{710}{\nano\meter}) (see Appendix \hyperref[sec:D]{D}). Figures \hyperref[fig:Figure2]{2(a)} and \hyperref[fig:Figure2]{2(b)} show typical Raman spectra recorded from Janus monolayer WSSe at room and cryogenic temperatures, respectively. From the comparison of Raman spectra at \SI{10}{\kelvin} [Fig. \hyperref[fig:Figure2]{2(b)}] with the calculated PhDOS in Fig. \hyperref[fig:Figure2]{2(c)}, where the dashed blue lines indicate the calculated values of  $\mathbf{\Gamma}$ phonons, we initially assign the first-order Raman modes $E^1$ at $\sim$\SI{207}{\centi\meter^{-1}} for $\lambda_{\mathrm{ex}}=\SI{633}{\nano\meter}$ ($\sim$\SI{209}{\centi\meter^{-1}} for $\lambda_{\mathrm{ex}}=\SI{532}{\nano\meter}$), $A_1^1$ at $\sim$\SI{285}{\centi\meter^{-1}} ($\sim$\SI{285}{\centi\meter^{-1}}), $E^2$ at $\sim$\SI{335}{\centi\meter^{-1}} ($\sim$\SI{336}{\centi\meter^{-1}}), and $A_1^2$ at $\sim$\SI{422}{\centi\meter^{-1}} ($\sim$\SI{423}{\centi\meter^{-1}}). 
All predicted first-order Raman modes, indicated in the spectra by blue arrows, are visible in all experimental conditions, albeit their intensity is maximum at 10 K and for $\lambda_{\mathrm{ex}} = \SI{633}{\nano\meter}$, which is close to the $A$ exciton resonance (top valence band to conduction band, see Appendix \hyperref[sec:D]{D}). 
The experimental results closely match the theoretical predictions. For the spectra acquired at $\lambda_{\mathrm{ex}}=\SI{633}{\nano\meter}$, we observe a broad background signal above \SI{\sim350}{\centi\meter^{-1}}, stemming from the photoluminescence tail of the material, due to the energetic proximity to the exciton transition. The Raman peaks $E^1$ at \SI{\sim204}{\centi\meter^{-1}} and $A_1^2$ at \SI{\sim420}{\centi\meter^{-1}} appear to be asymmetric, which can be attributed to phonon confinement effects due to an imperfect crystal quality \cite{Bersani.1998, Frey.1999}.
The experimental spectra also reveal a peak at $\SI{\sim152}{\centi\meter^{-1}}$, as indicated by gray arrows, which is especially strong at \SI{10}{\kelvin} and $\lambda_{\mathrm{ex}} = \SI{532}{\nano\meter}$, and corresponds to the position of the $A'^1$ mode in the LA branch at the \textbf{M} point or the $\prescript{1}{}E^{1}$ mode in the LA branch at the \textbf{K} point. These modes are expected to be silent in first-order Raman processes since their $|\textbf{q}|>0$. We can exclude that this peak is caused by higher-order Raman modes due to its low energy and, therefore,  attribute its appearance to defect-activation that relaxes the $\textbf{q}\approx\mathbf{0}$ selection rule \cite{Cancado.2011, Mignuzzi.2015}.

\begin{figure*}[ht!]
\centering
\includegraphics{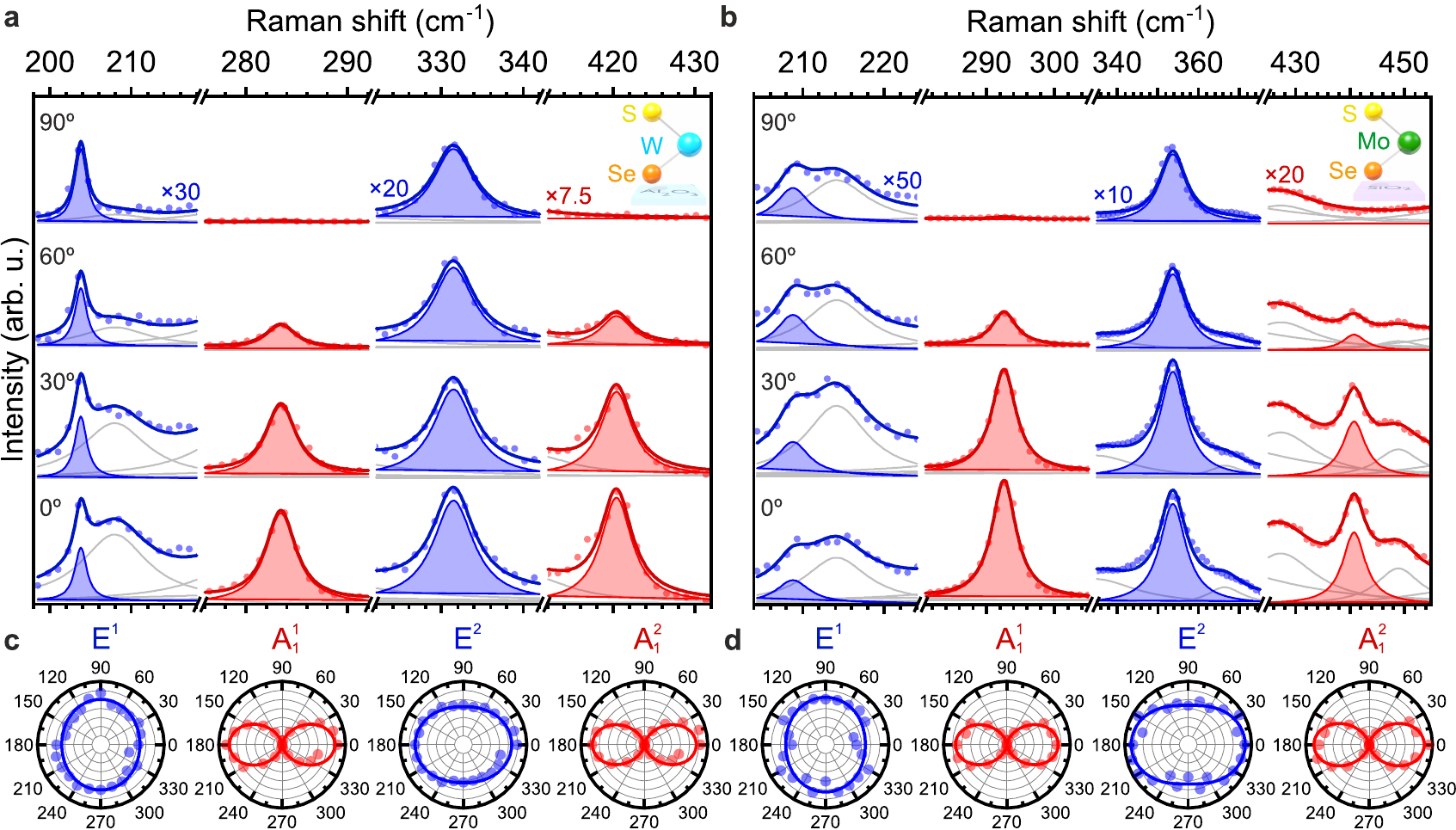}
\caption{\textbf{Polarization-resolved, first-order Raman modes.} Raman spectra of Janus monolayers \textbf{(a)} WSSe and \textbf{(b)} MoSSe for 0$^{\circ}$, 30$^{\circ}$, 60$^{\circ}$, and 90$^{\circ}$ angle between excitation and detection polarization, with experimental data shown as dots and their respective fits as solid curves; $E$ and $A_1$ Raman modes are plotted in blue and red, respectively. The spectra and fits are scaled with respect to the $A_1^1$ mode for clarity. Fits of the neighboring peaks are represented in gray. Polar plots of the integrated peak intensities of the complete polarization-resolved data set collected from \textbf{(c)} WSSe and \textbf{(d)} MoSSe as a function of the angle between excitation and detection polarization (dots), and their respective fits (solid curves). $E$ and $A_1$  modes are represented in blue and red, respectively. The spectra are taken at room temperature and $\lambda_{\mathrm{ex}}=\SI{532}{\nano\meter}$.}
\label{fig:Figure3}
\end{figure*}

The Raman spectra of Janus monolayer MoSSe are presented in Figs. \hyperref[fig:Figure2]{2(d)} and \hyperref[fig:Figure2]{2(e)}, accompanied by the calculated PhDOS in Fig. \hyperref[fig:Figure2]{2(f)}. Analogously to Janus monolayer WSSe, we compare the spectra to the theoretically predicted phonon energies [Fig. \hyperref[fig:Figure2]{2(f)}, dashed blue lines] and initially assign the first-order Raman modes $E^1$ at $\sim$\SI{212}{\centi\meter^{-1}} for $\lambda_{\mathrm{ex}}=\SI{633}{\nano\meter}$ ($\sim$\SI{211}{\centi\meter^{-1}} for $\lambda_{\mathrm{ex}}=\SI{532}{\nano\meter}$), $A_1^1$ at $\sim$\SI{291}{\centi\meter^{-1}} ($\sim$\SI{293}{\centi\meter^{-1}}), $E^2$ at $\sim$\SI{357}{\centi\meter^{-1}} ($\sim$\SI{356}{\centi\meter^{-1}}), and $A_1^2$ at $\sim$\SI{442}{\centi\meter^{-1}} ($\sim$\SI{444}{\centi\meter^{-1}}). 
Here, the spectra are strongly affected by the $\lambda_{\mathrm{ex}}$. First-order Raman modes, indicated by the blue arrows, are all clearly visible at $\lambda_{\mathrm{ex}}=\SI{633}{\nano\meter}$, while two peaks are much weaker at $\lambda_{\mathrm{ex}}=\SI{532}{\nano\meter}$. This arises once $\lambda_{\mathrm{ex}}=\SI{633}{\nano\meter}$ (\SI{1.96}{\electronvolt}) is close to the $B$ exciton transition (bottom valence band to conduction band, see Appendix \hyperref[sec:D]{D}), thereby increasing the Raman cross section.
Again, we observe good agreement between theory and experiment, well within the $\sim$\SI{8}{\centi\meter^{-1}} error of the calculations. Also for MoSSe, peaks appear around 155 and $\sim$\SI{175}{\centi\meter^{-1}} for $\lambda_{\mathrm{ex}} = \SI{633}{\nano\meter}$, as indicated by gray arrows, whose intensity is enhanced at 10 K. Similar to WSSe, the energy of the peaks in the $\sim$\SI{155}{\centi\meter^{-1}} range corresponds to the $A'^1$ mode in the TA branch at the \textbf{M} point and to the $A^1$ and $^2E^1$ modes in the TA and ZA branches at the \textbf{K} point. The peaks in the $\sim$\SI{175}{\centi\meter^{-1}} range correspond to the $A'^1$ mode in the LA branch at the \textbf{M} point and the $^1E^1$ mode at the \textbf{K} point in the LA branch. As discussed above, we exclude these peaks to be the result of higher-order Raman transitions due to their low energy, and instead attribute their appearance to defect activation. Interestingly, the presence of defect-activated Raman modes can be used to monitor the defect concentration in the crystal, in analogy to the \textit{D} peak in graphene \cite{Cancado.2011}, and as such constitutes precious information to assess the crystal quality. The peak at $\sim$\SI{272}{\centi\meter^{-1}} corresponds to the $^1E^2$ mode in the first ZO branch at \textbf{K}. However, due to the presence of other non-double-resonant phonon combinations matching the same energy, its assignment requires further investigation. For completeness, all first-order Raman modes at $\mathbf{\Gamma}$ are summarized in Table \ref{tab:Table1}.

\begin{table}[h!]
\begin{tabular}{|c|c|ccc|ccc|}
\hline
\multirow{2}{*}{$\lambda_{\mathrm{ex}}$} & \multirow{2}{*}{\begin{tabular}[c]{@{}c@{}}Phonon\\ mode\end{tabular}} & \multicolumn{3}{c|}{WSSe ($\mathrm{cm}^{-1}$)}                                       & \multicolumn{3}{c|}{MoSSe ($\mathrm{cm}^{-1}$)}                                      \\ \cline{3-8} 
                                          &                              & \multicolumn{1}{c|}{298 K} & \multicolumn{1}{c|}{10 K} & Theory & \multicolumn{1}{c|}{298 K} & \multicolumn{1}{c|}{10 K} & Theory \\ \hline\hline
\multirow{4}{*}{633 nm}                   & $E^1$                        & 204                        & 207                       & 204    & 210                        & 212                       & 208    \\
                                          & $A_1^1$                      & 282                        & 285                       & 282    & 290                        & 291                       & 293    \\
                                          & $E^2$                        & 333                        & 335                       & 331    & 354                        & 357                       & 358    \\
                                          & $A_1^2$                      & 420                        & 422                       & 420    & 439                        & 442                       & 445    \\ \hline
\multirow{4}{*}{532 nm}                   & $E^1$                        & 204                        & 209                       & 204    & 209                        & 211                       & 208    \\
                                          & $A_1^1$                      & 284                        & 285                       & 282    & 292                        & 293                       & 293    \\
                                          & $E^2$                        & 331                        & 336                       & 331    & 354                        & 356                       & 358    \\
                                          & $A_1^2$                      & 420                        & 423                       & 420    & 441                        & 444                       & 445    \\ \hline
\end{tabular}
\caption{\textbf{List of first-order Raman modes of Janus monolayers WSSe and MoSSe at all measured experimental conditions and comparison with the DFPT calculations.}
}
\label{tab:Table1}
\end{table}

To further confirm the peaks’ assignments, we performed linear polarization-dependent Raman scattering measurements at room temperature and $\lambda_{\mathrm{ex}}=\SI{532}{\nano\meter}$ (see Appendix \hyperref[sec:A]{A} for details). Figures \hyperref[fig:Figure3]{3(a)} and \hyperref[fig:Figure3]{3(b)} show the Raman spectra of Janus monolayers WSSe and MoSSe, respectively, for four representative angles between the excitation and detection polarization, starting from 0$^{\circ}$ (copolarized configuration) to 90$^{\circ}$ (cross-polarized configuration). The experimental data (dots) are fitted with a sum of Lorentzians (blue and red thick curves). Going from co- to cross polarization, the intensity of the blue peaks (initially assigned as $E$ modes) remains overall constant, while the intensity of the red peaks (initially assigned as $A_1$ modes) decreases from the maximum at 0$^{\circ}$ to zero at 90$^{\circ}$. Figures \hyperref[fig:Figure3]{3(c)} and \hyperref[fig:Figure3]{3(d)} show the polar plots of the fitted peak intensities of Janus monolayers WSSe and MoSSe as a function of the angle between excitation and detection linear polarizer for the complete data set acquired (see Appendix \hyperref[sec:E]{E} for all raw spectra). Due to their symmetry, Raman selection rules predict the intensity of the $E$ modes $I_E$ to be independent from the linear polarization angle as $I_{E^1,E^2}  = c^2$, while the intensities of $A_1$ modes $I_{A_1}$ are predicted to be polarized following $I_{A_1^1,A_1^2}  = a^2\cos^{2}\theta$, where $c$ and $a$ are constants and $\theta$ is the angle. Thus, linearly polarized Raman spectra confirm our initial peaks’ assignments.

\begin{figure*}[ht!]
\centering
\includegraphics{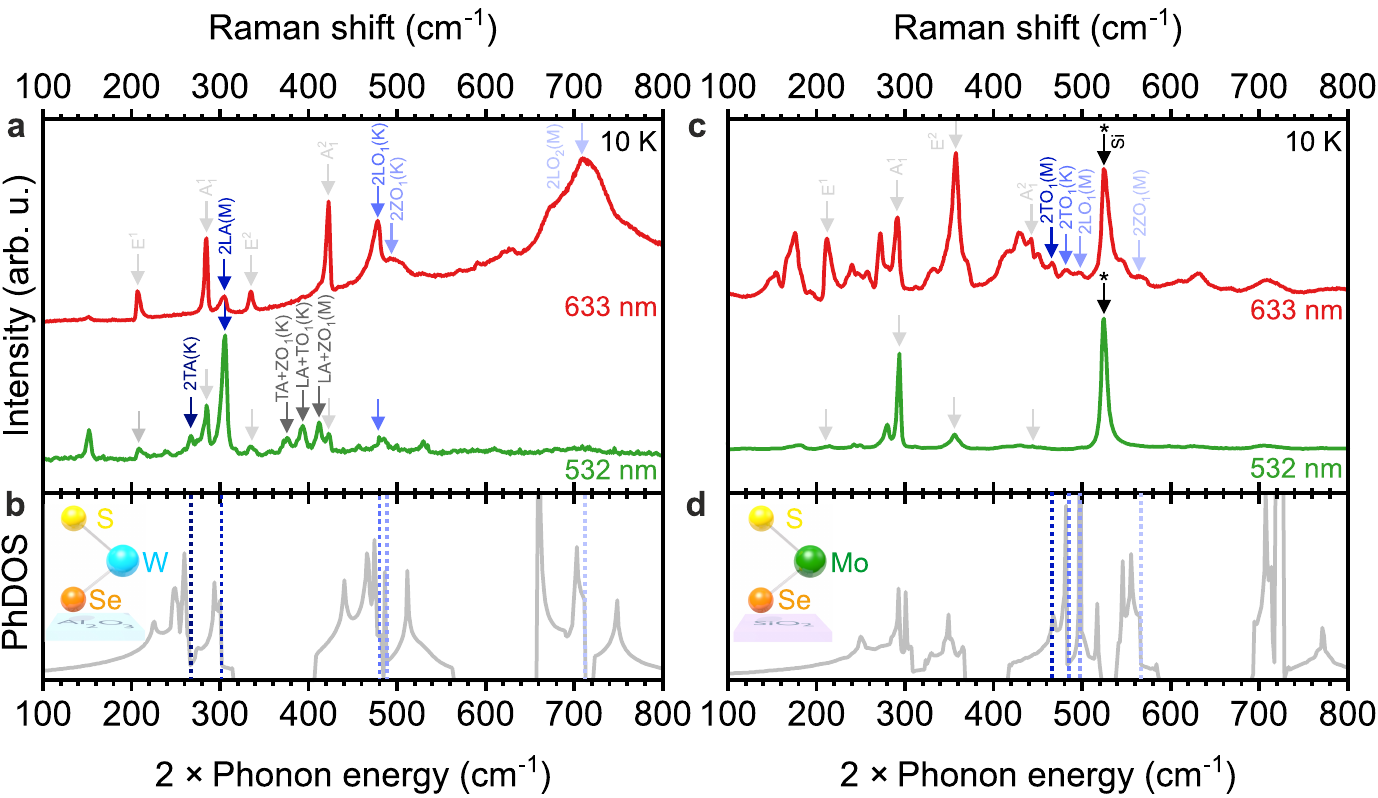}
\caption{\textbf{Higher-order Raman processes in Janus WSSe and MoSSe monolayers.} Raman spectra of \textbf{(a)} WSSe and \textbf{(c)} MoSSe at $\lambda_{\mathrm{ex}}=\SI{532}{\nano\meter}$ (green curves) and $\lambda_{\mathrm{ex}}=\SI{633}{\nano\meter}$ (red curves). Blue-shaded arrows indicate Raman peaks corresponding to double-resonant processes with mode assignments following Table \ref{tab:Table2}. Defect-assisted processes are indicated by dark gray. For comparison, light gray denotes first-order modes. All measurements are taken at 10 K. Black arrows indicate the Raman signal of the substrate. PhDOS of double-resonant phonon modes of \textbf{(b)} WSSe and \textbf{(d)} MoSSe. Dotted lines indicate energies of double-resonant phonons that correspond to observed Raman peaks indicated by the same color.}
\label{fig:Figure4}
\end{figure*}

To explore higher-order Raman peaks, in Figs. \hyperref[fig:Figure4]{4(a)} and \hyperref[fig:Figure4]{4(c)} we plot the Raman spectra of Janus monolayers WSSe and MoSSe at 10 K between 100 and \SI{800}{\centi\meter^{-1}}, and compare them to the PhDOS as a function of twice the phonon energy in Figs. \hyperref[fig:Figure4]{4(b)} and \hyperref[fig:Figure4]{4(d)}. This is motivated by the role of double-resonant Raman scattering \cite{Thomsen.2000} in higher-order Raman transitions. In double-resonant Raman scattering, two phonons with the same momentum but opposite direction make electrons scatter far from their excitation point in the Brillouin zone and then come back to the initial position, through two resonant and two nonresonant scattering events, satisfying $\textbf{q}\approx\mathbf{0}$. Double-resonant Raman processes in WSSe and MoSSe are indicated by blue arrows in Figs. \hyperref[fig:Figure4]{4(a)} and \hyperref[fig:Figure4]{4(c)}. Higher-order scattering processes that include phonons from different branches are also energetically allowed through defect activation that locally breaks crystal symmetries (dark gray arrows). A comprehensive assignment list of the observed higher-order Raman peaks is given in Table \ref{tab:Table2}. The unassigned peaks which do not match with double-resonant processes may be defect activated, however, further studies are required to elucidate their nature.

\begin{table}[h!]
\begin{tabular}{|c|c|ccc}
\cline{1-2} \cline{4-5}
\multicolumn{2}{|c|}{WSSe}     & \multicolumn{1}{c|}{} & \multicolumn{2}{c|}{MoSSe}                                 \\ \cline{1-2} \cline{4-5} 
Peak (cm$^{-1}$) & Assignments & \multicolumn{1}{c|}{} & \multicolumn{1}{c|}{Peak (cm$^{-1}$)} & \multicolumn{1}{c|}{Assignments} \\ \hhline{==~==}
$267^{\dagger}$ & 2TA(K)       & \multicolumn{1}{c|}{} & \multicolumn{1}{c|}{466} & \multicolumn{1}{c|}{2TO$_1$(M)} \\ 
305             & 2LA(M)       & \multicolumn{1}{c|}{} & \multicolumn{1}{c|}{482} & \multicolumn{1}{c|}{2TO$_1$(K)} \\ 
$376^{\dagger}$ & TA+ZO$_1$(K) & \multicolumn{1}{c|}{} & \multicolumn{1}{c|}{497} & \multicolumn{1}{c|}{2LO$_1$(M)} \\
$394^{\dagger}$ & LA+TO$_1$(K) & \multicolumn{1}{c|}{} & \multicolumn{1}{c|}{564} & \multicolumn{1}{c|}{2ZO$_1$(M)} \\ \cline{4-5}
$412^{\dagger}$ & LA+ZO$_1$(M) &                       &                          &                                 \\ 
478             & 2LO$_1$(K)   &                       &                          &                                 \\
491             & 2ZO$_1$(K)   &                       &                          &                                 \\ 
709             & 2LO$_2$(M)   &                       &                          &                                 \\ \cline{1-2}
\end{tabular}
\caption{\textbf{List of higher-order Raman modes of Janus monolayers WSSe and MoSSe, listed by experimental energy from spectra collected at $\lambda_{\mathrm{ex}} = \SI{633}{\nano\meter}$ ( $^{\dagger}\lambda_{\mathrm{ex}} = \SI{523}{\nano\meter}$) and 10 K.}} 
\label{tab:Table2}
\end{table}


\section{Conclusion} 

In summary, we presented a combined theoretical and experimental study of the Raman modes of Janus monolayers WSSe and MoSSe, whereby we found excellent agreement between the two for the frequencies of first- and higher-order modes. Moreover, we discovered the presence of defect activation of otherwise silent Raman modes, which may be used as markers for assessing crystal quality in further studies. 

The recent synthesis of Janus monolayer TMDs adds an extra degree of freedom to the wide family of two-dimensional and layered materials, with the potential for tunable, strongly interacting dipolar excitons in single-layer materials as the stepping stone for further exploration of correlated many-body states, exciton transport, and applications that exploit such features. However, novel physics and exciting new applications require in-depth information over the materials' properties and growth quality, with Raman spectroscopy being a widely utilized technique in a such regard due to its descriptive power and simplicity of use and interpretation. Our work sets a general and much-needed reference over the vibrational properties of Janus monolayers, provides a starting point for further investigations on the role of phonons in such materials, and enables the benchmarking of future crystal growth attempts. 


\begin{acknowledgments}
M.M.P., M.K., and M.B. contributed equally to this work. M.B., J.J.F., A.R.B.-M., and K.M. conceived and managed the project. Y.Q., Y.Sa., Y.Sh., and S.T. grew the Janus TMD monolayers. M.M.P., M.K., and M.B. performed the optical measurements and analyzed the data. A.R.B.-M. performed DFPT calculations. All authors participated in the discussion of the results and the writing of the manuscript.

We thank Moritz Meyer for technical assistance. S.T. acknowledges support from NSF DMR-1955889, NSF CMMI-1933214, NSF DMR-1552220, and DOE-SC0020653. K.M. and J.J.F. acknowledge support from the European Union Horizon 2020 research and innovation programme under Grant Agreement No. 820423 (S2QUIP) and the Deutsche Forschungsgemeinschaft (DFG, German Research Foundation) under Germany’s Excellence Strategy – MCQST (EXC-2111) and e-Conversion (EXC-2089). M.M.P. acknowledges TUM International Graduate School of Science and Engineering (IGSSE). M.K. acknowledges support from the International Max Planck Research School for Quantum Science and Technology (IMPRS-QST). M.B. acknowledges support from the Alexander von Humboldt Foundation. K.M. acknowledges support from the Bayerische Akademie der Wissenschaften. A.R.B.-M. acknowledges support from  DGTIC-UNAM Supercomputing Center under Project LANCAD-UNAM-DGTIC-359.

\end{acknowledgments}



\appendix
\section{METHODS}
\label{sec:A}

\subsection{DFPT}
Phonon dispersion relations were calculated using density functional perturbation theory with the local-density approximation to the exchange-correlation function \cite{Giannozzi.2017, Sohier.2017}. The vacuum distance between neighboring layers was \SI{20}{\angstrom} to describe isolated layers within the periodic boundary conditions. Norm-conserving pseudopotentials and a basis set defined from a energy cutoff of 105 Ry \cite{Hamann.2013, vanSetten.2018} were used. The first Brillouin zone was sampled with a 15$\times$15$\times$1 Monkhorst-Pack grid.
\subsection{SEAR}
The synthesis of Janus TMDs is carried out in a specially designed quartz chamber and a home-built inductively coupled plasma system. The plasma chamber consists of a 5-ft-long quartz tube with a 1-in. inner diameter suspended off-centered on a Lindberg Blue/M single-zone furnace. A copper coil with a length  $\sim$1.5 in. consisting of about five turns were wound around the quartz tube. The end of the Cu coil is connected to a 100-W tunable rf source (SEREN R101) through a custom-designed impedance match network. One end of the quartz tube is connected to an Edwards vacuum pump while the other end is fitted with a hydrogen supply line. Hydrogen flow rate and the pressure within the chamber is regulated by means of a capacitance manometer and a pressure controller. For the synthesis of Janus TMDs, the reaction chamber is pumped down to a base pressure of 15 mTorr, after which the chamber was purged with 20-sccm H$_2$ flow, maintaining an operation pressure at 300 mTorr. The plasma was generated with 15-W rf power, and the visible plasma tail position was marked on the quartz tube. For the SEAR process to create WSSe, chemical vapor deposition (CVD)-grown WSe$_2$ was placed 4 cm upstream of the marked visible plasma tail position onto a quartz boat, and 2-g sulfur was placed 15 cm upstream of the H$_2$ plasma tail. The plasma treatment lasted for 18 min. For the creation of MoSSe, the position of CVD-grown MoSe$_2$ and S source were kept at the same position as WSSe, except the processing time was decreased to 8 min because of the lower Mo-Se bond energy. The SEAR process can also be set up to create a 2D Janus structure from sulfur-based TMDs and selenium precursors in a similar fashion by varying the processing parameters.

\subsection{Optical setup}
Raman and photoluminescence (PL) measurements were done using a custom-made confocal microscope in backscattering geometry. The excitation laser beam is focused on the sample by an objective with numerical aperture NA $=$ 0.75 to a diffraction-limited spot. For cryogenic measurements a He-flow cryostat (Cryovac) was used. The collected light is analyzed in a spectrometer (maximum point-to-point resolution $\sim$1.2 cm$^{-1}$ for  $\lambda_{\mathrm{ex}}=\SI{532}{\nano\meter}$ and $\sim$0.8 cm$^{-1}$ for  $\lambda_{\mathrm{ex}}=\SI{633}{\nano\meter}$ with a grating with 1200 lines/mm) coupled to a charged-coupled device (Horiba). Both incident ($\mathbf{e}_i$) and scattered ($\mathbf{e}_s$) light polarization vectors are placed in the $xy$ plane, which is relevant to account for the Raman intensity  $I = |\mathbf{e}_s R\mathbf{e}_i|^2$ (for the case of no polarization filtering in the detection path, $I \propto |\mathbf{e}_x R\mathbf{e}_i|^2 + |\mathbf{e}_y R\mathbf{e}_i|^2$) \cite{Saito.2016}. Here, $R$ represents the Raman tensor, which is derived from the point group symmetry, and alongside the scattering configuration dictates the selection rules \cite{Loudon.2001, Kuzmany.2009}. For polarization-resolved Raman measurements, we excited the Janus crystals using fixed linearly polarized light at $\lambda_{\mathrm{ex}}=\SI{532}{\nano\meter}$ and room temperature, collecting the scattered light through a rotating half-waveplate and a fixed linear polarizer.

\section{JANUS TMDs IN $k$ SPACE}
\label{sec:B}
\begin{figure}[h]
\centering
\includegraphics[scale=0.5]{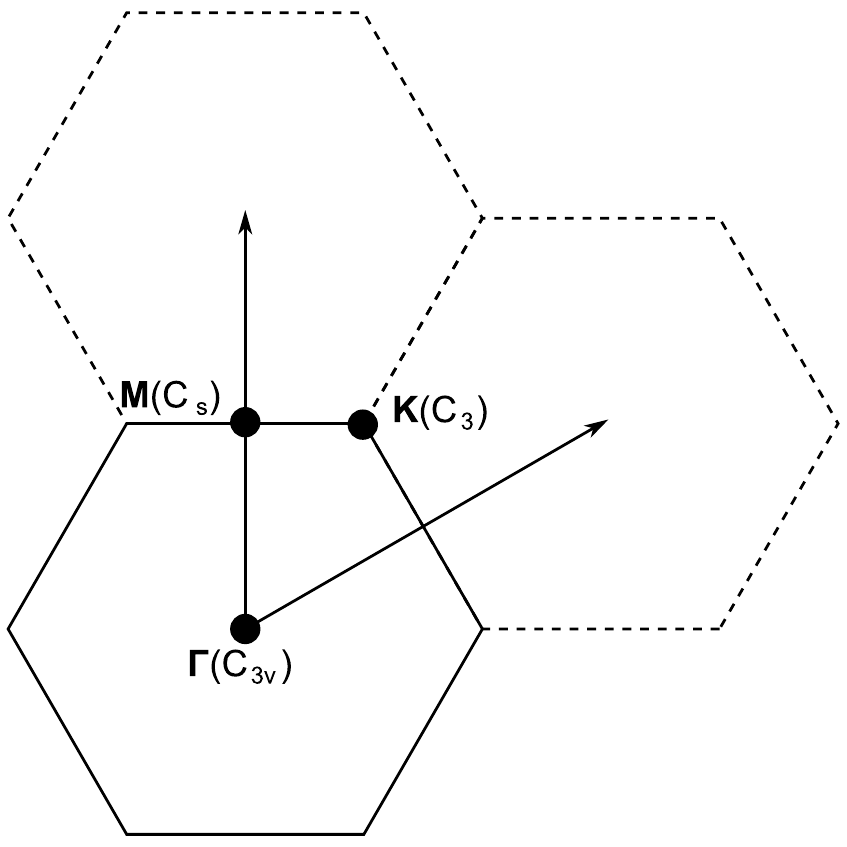}
\caption{\textbf{Groups and subgroups for the high-symmetry points of the Brillouin zone of Janus TMDs.}}
\label{fig:Figure5}
\end{figure}

Figure \ref{fig:Figure5} shows the Brillouin zone of the Janus TMDs. The zone center $\mathbf{\Gamma}$ point has the same symmetry as the crystal, while the $\mathbf{K}$ and $\mathbf{M}$ high-symmetry points are subgroups with $C_3$ and $C_s$ symmetries, respectively.

\section{PHONON BAND GAPS}
\label{sec:C}
The phonon band structure of Janus TMD monolayers [Figs. \hyperref[fig:Figure1]{1(b)} and \hyperref[fig:Figure1]{1(c)}, main text] reveals the gap of unaccessible phonons that separates acoustic from optical bands. WSSe and MoSSe exhibit $\Delta^{\mathrm{WSSe}}_{\mathrm{TO_1-LA}}\approx\SI{46}{cm^{-1}}$ and $\Delta^{\mathrm{MoSSe}}_{\mathrm{TO_1-LA}} \approx \SI{24}{cm^{-1}}$ band gaps, respectively. Forbidden phonon energies occur also between optical bands (phonon band gaps of $\Delta^{\mathrm{WSSe}}_{\mathrm{TO_2-ZO_1}} \approx \SI{46}{cm^{-1}}$ for WSSe and $\Delta^{\mathrm{MoSSe}}_{\mathrm{TO_2-ZO_1}} \approx \SI{53}{cm^{-1}}$ for MoSSe). 

\section{PHOTOLUMINESCENCE OF WSS\lowercase{e} AND M\lowercase{o}SS\lowercase{e}}
\label{sec:D}

\begin{figure}[h]
\centering
\includegraphics[scale=0.8]{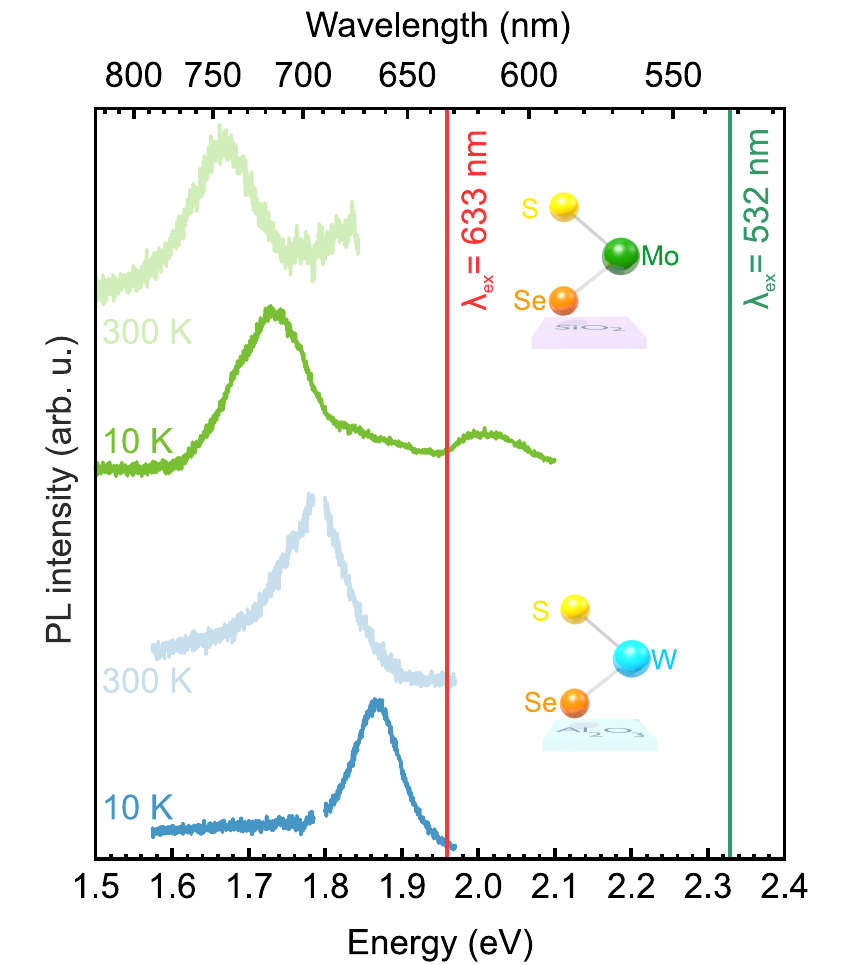}
\caption{\textbf{Photoluminescence spectra of Janus WSSe and MoSSe monolayers at 10 and 300 K.} The laser excitation wavelength $\lambda_{\mathrm{ex}}=\SI{633}{\nano\meter}$ is close to the \textit{A} exciton resonance in WSSe (dark blue) and to the \textit{B} exciton resonance in MoSSe (dark green).}
\label{fig:Figure6}
\end{figure}

\section{POLARIZATION-RESOLVED RAMAN SPECTRA}
\label{sec:E}

\begin{figure}[h]
\centering
\includegraphics{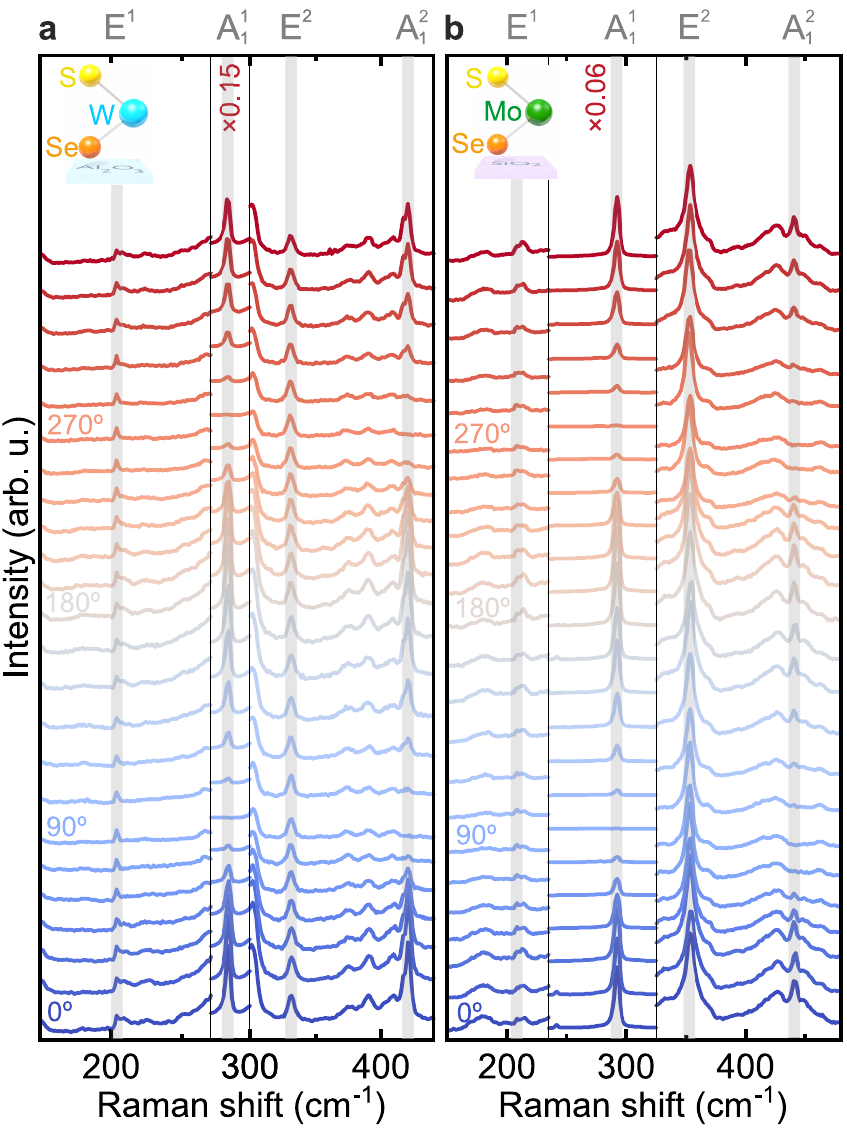}
\caption{\textbf{Complete set of the polarization-resolved spectra of Janus TMD monolayers.} Specific peaks are rescaled as indicated in the figure for clarity.}
\label{fig:Figure7}
\end{figure}

\end{document}